\documentclass[aps,prd,twocolumn]{revtex4}
\usepackage{amsmath}
\usepackage{graphicx}
\usepackage[colorlinks=true,urlcolor=blue,linkcolor=blue]{hyperref}

\newcommand{\mt}{{\tilde m}}
\newcommand{\Ft}{{\tilde F}}
\newcommand{\Mt}{{\tilde M}}
\newcommand{\chit}{{\tilde \chi}}
\newcommand{\Jt}{{\tilde J}}

\newcommand{\dth}{\Delta \omega}
\newcommand{\dch}{\Delta \chi}

\newcommand{\dthp}{\left( \Delta \omega \right)}

\newcommand{\lx}{\lambda}

\newcommand{\ex}{\epsilon}
\newcommand{\be}{\begin{equation}}
\newcommand{\ee}{\end{equation}}

\def \gta {\mathrel{\vcenter
     {\hbox{$>$}\nointerlineskip\hbox{$\sim$}}}}
\begin{document}

\title{Localization of Gauge Fields and Monopole Tunnelling}
\author{G. Dvali$^{1,2}$, H.B. Nielsen$^{1,3}$ and N. Tetradis$^4$ 
}\date{\today}

\begin{abstract}
We study the dynamical localization of a massless gauge field 
on a lower-dimensional surface (2-brane). 
In flat space, the necessary and sufficient condition for this phenomenon is the 
existence of confinement in the bulk. 
The resulting configuration is equivalent to a dual Josephson junction. 
This duality leads to an interesting puzzle, as it implies that a localized 
massless theory, even in the Abelian case, must become confining at 
exponentially large distances.
Through the use of topological arguments we clarify the physics behind this 
large-distance confinement and identify the  
instantons of the brane world-volume theory that are responsible for its appearance. 
We show that they correspond to the (condensed) 
bulk magnetic charges (monopoles), that occasionally tunnel through 
the brane and induce weak confinement of the brane theory. 
We consider the possible generalization of this effect to higher dimensions
and discuss phenomenological bounds on the  
confinement of electric charges at exponentially large distances 
within our Universe.

\end{abstract}

\affiliation{
$^1$
CERN, Theory Division, CH-1211 Geneva 23, Switzerland
\\
$^2$
Center for Cosmology and Particle Physics, Department of Physics, New York University, 
New York NY 10003, USA
\\
$^3$
The Niels Bohr Institute, Copenhagen DK 2100, Denmark
\\
$^4$
Department of Physics, University of Athens, Zographou 157 84, Athens, Greece}
\clearpage
\maketitle
\section{Introduction}

The dynamical localization of massless modes is a very interesting physical phenomenon,
that goes against the naive quantum-mechanical intuition according to which a bound state 
naturally has a mass  of the order of the inverse  localization width.   The exceptions 
from this  ``rule''  are  well known for spin-$0$  and spin-${1}/{2}$ systems. The massless 
spin-0 scalars can be localized  on lower-dimensional solitons  
as Goldstone bosons (sound waves) of the broken translational invariance.  
The appearance of the fermionic zero modes in the field of topological defects 
is due to topologically non-trivial boundary conditions for the fermion mass and is guaranteed 
by the index theorem \cite{index}.

In contrast, no analogue of the Goldstone or the index theorems exists for the dynamical localization  
of massless spin-$1$ fields.  In string theory the objects that support massless spin-1 excitations 
on their world volume are $D$-branes \cite{polchinski}. 
The massless gauge fields are excitations of the open strings  
that end on the brane. Therefore, they are intrinsically  lower-dimensional ``from the beginning''. 
In other 
words,  it is unclear  how to trace their higher-dimensional counterparts.  In the present work we shall be 
concerned with the dynamical localization of gauge fields within field theory. 

It was argued in \cite{giamisha} that,  at least on  asymptotically flat spaces,  the necessary and 
 sufficient condition for the localization of massless gauge fields on a lower-dimensional surface, 
 embedded in a higher-dimensional bulk space, is {\it bulk confinement}. That is, at high energies 
 the localized gauge theory must become part of a confining higher-dimensional theory.  
As shown in \cite{giamisha}, a  simple 
$(3+1)\,  \rightarrow \, (2+1)$  model illustrating such a dynamical localization of a massless photon 
is a $(3+1)$-dimensional $SU(2)$ gauge theory Higgsed  down to $U(1)$ on a $(2+1)$-dimensional 
wall (brane).  The dynamical localization of a (perturbatively) massless photon follows from the  
fact that the $U(1)$ gauge field becomes part of a confining gauge theory in the bulk, with a mass gap 
$\sim \Lambda$. A localized photon can only escape into the bulk in the form of a massive glueball.   
 Thus, the bulk confinement creates a barrier that confines the photon to the brane. 
The localized photon is 
 perturbatively massless because the $U(1)$ gauge symmetry is never Higgsed.   

  The immediate consequence of this picture is the existence of electric flux tubes that end on the 
  brane.  The close resemblance  between these and the open strings of the $D$-brane theory is 
  rather striking. It poses the question whether the connection between the open strings and 
  the massless gauge fields 
has a common underlying origin both in string and field theory \cite{dv}. 
  This analogy was also discussed earlier in the context of the connection between the domain walls of 
supersymmetric gauge theories and $D$-branes 
\cite{dbranes}.   
  
In the present work, we shall be concerned with phenomena that arise when one tries 
to think about the localization mechanism  of \cite{giamisha} as the creation of
a dual Josepson junction \cite{tetradis}. 
Indeed, the appearance of confinement is believed to be
equivalent to the condensation of magnetic charges \cite{dualsup}.  
The brane then becomes a dual insulator, sandwiched  
between two infinite magnetic superconductors. For the ordinary Josephson junction,
it is well known that the Meissner effect (the confinement of magnetic charges)  
penetrates the layer at the quantum-mechanical level, because of the presence of
tunnelling Josephson currents \cite{joseph}. 
Hence, even on the layer  the magnetic charges are not in a truly Coulomb 
phase,  but get confined  at large distances.  Because of the tunnelling nature of the effect, the 
magnetic confinement scale of the $(2+1)$-dimensional theory is exponentially suppressed 
by the width of the layer. Nevertheless, this is a real effect that has been observed. 
In complete analogy with the standard Josephson junction, 
the presence of tunnelling magnetic currents in the dual picture
is expected to induce exponentially weak 
confinement of electric charges located on the brane. 
We emphasize that the appearance of
a very small mass gap through confinement 
should not be confused with the presence of a mass term of the Higgs or Proca type for the photon.
Perturbatively the photon is massless, and only at exponentially large distances 
confinement sets in.

While trying to apply this reasoning to the  $SU(2)$ model of \cite{giamisha} one encounters 
a puzzle. As suggested in \cite{tetradis},   
if duality works the $(2+1)$-dimensional  $U(1)$ theory must become confining at exponentially large 
distances.  
On the other hand, one may wonder how this is possible, since at distances above the brane width  
an effective 
 $(2+1)$-dimensional theory is just a pure perturbatively massless  $U(1)$. Where is the 
confining dynamics coming from? 

In the present work we shall resolve the above  puzzle. We shall argue that the confinement 
of the effective
$(2+1)$-dimensional theory on the brane can be understood in terms of the confining dynamics of 
a truly $(2+1)$-dimensional theory with compact $U(1)$. 
The latter was first studied by Polyakov 
\cite{polyakov}. Perturbatively, the theory seems to be in the Coulomb phase. However, Polyakov
demonstrated that the IR dynamics is governed by instantons, whose 
presence results in the confinement of electric charges.  

The identification of the two low-energy theories requires some care.  
In order to achieve it we employ a topological method, in which we view a $D$-dimensional theory 
as a slice of a $(D+1)$-dimensional one.  This method  allows us to classify  the instantons 
of the $D$-dimensional theory as the monopoles of the  $(D+1)$-dimensional 
one that tunnel across the 
slice.  Applying this method to the model of \cite{giamisha}, we show that the
instantons  are the monopoles  that tunnel through the brane.  These instantons lead to the 
confinement of electric charges at exponentially large scales, in agreement with the conclusions 
of \cite{tetradis}.

We shall also discuss briefly the relevance of this phenomenon for higher-dimensional cases, 
and possible  phenomenological implications of the large-distance electromagnetic confinement. 
We point out that the existence of the galactic magnetic field already puts a very severe restriction 
on the confinement scale. This is different to the situation 
for a photon mass of the Higgs or Proca type, which 
essentially is not restricted by the galactic magnetic field \cite{adg}.

\section{Localization of gauge fields}  
\label{localization}
\subsection{Problems with Localization by the Bulk Higgs Effect} 

If the bulk mass that confines the 
photon to a lower-dimensional sub-surface or layer (``brane'') is of the Higgs or Proca 
type, the localization does not produce a massless
electric field in the effective lower-dimensional theory. 
The localized electric photon acquires a large mass because of the same bulk Higgs effect that confines 
it to the brane. 
As the gauge field is in the Higgs phase in the bulk, electric charge screening also penetrates  
the brane
resulting in an exponentially decaying electric field. 
The problem can be illustrated by the following simple model that localizes 
the photon onto a $(2+1)$-dimensional surface. Consider an Abelian Higgs model
and assume a potential for the Higgs field with two degenerate minima: one located at $\Phi=0$ and the other 
at $|\Phi| = M \neq 0$. In the $\Phi=0$ vacuum the photon is massless and the $U(1)$-theory is in the 
Coulomb phase. The test charges create  a long-range $1/r$ potential.   In the $\Phi \neq 0$ vacuum 
the photon has a mass $m_{\gamma} \sim M$ and the theory is in the Higgs phase.  
In this vacuum the test charges are screened and generate a Yukawa-type potential 
$\sim {\exp({-m_{\gamma}r}) / r}$.  

The two phases can coexist. In such a case they will be separated by domain 
walls of tension $\sim M^3$ and width $\sim M^{-1}$.  
Throughout the wall the expectation value of $|\Phi|$ 
smoothly  interpolates between $0$ to $M$ over a distance $\sim M^{-1}$. 
One simple choice of $V(\Phi)$ that realizes this picture is  
  \begin{equation}
\label{potu1}
V(\Phi) \, =  \, (|\Phi|^2 \, - \, M^2)^2 {\frac{|\Phi|^2}{M^2}}.
\end{equation}  
(For simplicity we have set all the dimensionless coupling constants equal to one). 
This theory has a single mass scale $M$ and vacua at 
$\Phi =0$ and $ |\Phi|= M$. We have chosen the parameters in such a way that the vacua are degenerate.  
This choice allows for the two phases to coexist, with a static wall in between. 
Of course, many other choices of $V(\Phi)$ with similar properties are possible \cite{dv}. 
 
Consider a situation in which a layer of the would-be Coulomb vacuum with $\Phi=0$ is ``trapped'' 
between two $|\Phi| = M$ phases. This can be achieved by placing a wall and  
an anti-wall parallel to each other at a certain 
distance $d$.  If $z$ is the coordinate perpendicular to the walls,  
$|\Phi|$ would vary from $M$ to $0$ and back to $M$, as one crosses the wall and anti-wall 
system  from $z =-\infty$ to $z= + \infty$.   
As we said above, the width of the walls  that bound the $\Phi=0$ vacuum is $\sim M^{-1}$. 
Of course, this system is not really stable, and not even static, 
but if $d \gg M^{-1}$ the force between the wall and anti-wall is exponentially suppressed by a factor 
$\exp({-Md})$, so that the system can be considered to be static over the time-scales of interest. 
    
Naively, one may think that this setup localizes a massless photon in an effective 
$(2+1)$-dimensional theory on the layer, because $\Phi = 0$ there. 
However, this is not the case because
the would-be Coulomb vacuum is sandwiched between two Higgs phases
and the charge screening penetrates there. 
Formally, the absence of the massless mode can be seen directly from the 
mode expansion analysis, but the physical reason is clear: The two Higgs phases bounding the 
would-be Coulomb vacuum are superconductors, with a lot of free charges available in the vacuum. 
Any test charge placed in the layer polarizes this vacuum and creates image charges that screen it.  
In the language of electric flux lines, 
the fact that the bulk is superconducting implies that the electric flux lines of the 
$(2+1)$-dimensional source can end on the boundary in the vicinity of the image charges 
(see Fig. (\ref{josephson})). 
Hence, there is no flux conservation in the $(2+1)$-dimensional theory.   
The theory is in the Higgs phase, both on the layer and in the bulk, with comparable masses for the photon.  

\subsection{Localization by the Dual Higgs effect}    
The lesson from the previous analysis is that,  
in order to localize a massless photon, the electric flux lines
should not be able to either spread out of the brane or end on its boundary.   
In other words, the bulk medium must repel the electric flux lines, without breaking or terminating them 
on any image charges. This will induce flux conservation within the $(2+1)$-dimensional theory on the 
layer, and, by Gauss' law, the $(2+1)$-dimensional Coulomb phase. This means that 
the bulk condensate that repels the flux lines must be of the magnetic type.  
In such a case the electric charges in the bulk will be confined, but not screened. 
  
This can be achieved if the Abelian $U(1)$ symmetry becomes part of a confining theory
at a certain scale $\Lambda$.  The simple model (a variation of the original one) that realizes this 
consists of an $SU(2)$ gauge theory     
with a single adjoint  Higgs $\Phi^a$ ($a=1,2,3$). The potential is a  
non-Abelian generalization of (\ref{potu1}):
\begin{equation}
\label{potu2}
V(\Phi) \, =  \, (\Phi^a\Phi^a \, - \, M^2)^2 {\frac{\Phi^a\Phi^a}{ M^2}}. 
\end{equation}  
Again, this potential has the two vacua,   $\Phi^a = 0$ and   $\Phi^a = Mn^a$ 
(where $n^a$ is an arbitrary unit vector in three-dimensional space).   
In the first vacuum, the theory is in the confining phase and there is a mass gap $\sim \Lambda$. 
All the states in this vacuum are massive glueballs. 
In the second vacuum, the theory is Higgsed down to $U(1)$ and there is a massless photon.  
Perturbatively, test charges in this vacuum are in the Coulomb phase.  
As in the previous  example, the two phases may coexist,
separated by domain walls of tension $\sim M^3$ and width 
$\sim M^{-1}$.   

Let us now consider a layer of the Abelian $U(1)$ vacuum sandwiched between the two 
$SU(2)$ confining phases.  The width of the layer (distance between the wall and anti-wall) 
is $d$. We shall assume that   $M \gg d^{-1}, \Lambda$.  Again, the walls can be considered 
as being static in the time scales of interest. 
In the layer of the $U(1)$ vacuum all the states except the 
photon are massive. Although the photon has no Higgs or Proca mass anywhere, it is repelled out of 
the bulk by the confining dynamics. It cannot enter the bulk without 
becoming a massive glueball.  In other words, the electric flux of a test source can only penetrate 
the bulk in the form of an $SU(2)$ flux tube, with a tension $\sim \Lambda^2$.   
As a result, perturbatively
the effective low-energy theory below the scale $d^{-1}$ is a $(2+1)$-dimensional theory of a 
massless photon.  Since below such energies all the heavy states decouple, 
and there are no light charges available, 
one expects the photon to remain in the Coulomb phase down to arbitrarily low energies. 
But is this true? 

\subsection{The Puzzle} 

As far as the low-energy theory is concerned, the bulk confinement picture is dual to 
the bulk Higgs model.  In the $U(1)$ Higgs phase, test magnetic monopoles are 
confined because they are connected by the Nielsen-Olesen magnetic flux tubes \cite{vortex}. 
These flux tubes are dual to the ones that confine test charges in the unbroken $SU(2)$ phase. 
If duality holds, and the bulk confinement picture indeed creates an 
electric Coulomb phase on the layer, 
the bulk Higgs theory should create a magnetic Coulomb phase there.  
  
This analogy is precisely the source of the puzzle.  The crucial point is that in the bulk Higgs picture 
the bulk is a superconductor, and there must be a Josephson effect.  Because of the phase difference
of the condensates on the two sides of the layer, charges can tunnel through it and create a  
current. 
As a result, the 
magnetic charges on the layer cannot remain in  the Coulomb phase for arbitrarily large distances, but only
up to an exponentially large separation, after which they get confined.  In other words, the flux of 
the Nielsen-Olesen tube can only spread out to an exponentially large distance but not to infinity. 
This fact poses the question of whether a dual analogue of the Josephson effect takes place in the 
bulk confinement 
model, and whether the electric charges on the layer  remain in the Coulomb phase or become confined at 
exponentially large distances.

\section{The Physics of the Josephson Junction}

In this section we summarize the physics of the Josephson junction and discuss the
implications for the dual picture in which there is a condensate of magnetic charge
in the bulk. Our arguments are quite general and do not rely on the details of
the theory in which electric and magnetic charge are incorporated consistently.

\subsection{The Josephson Junction}

\begin{figure}[t]
\begin{center}
\includegraphics[clip,width=0.9\linewidth]{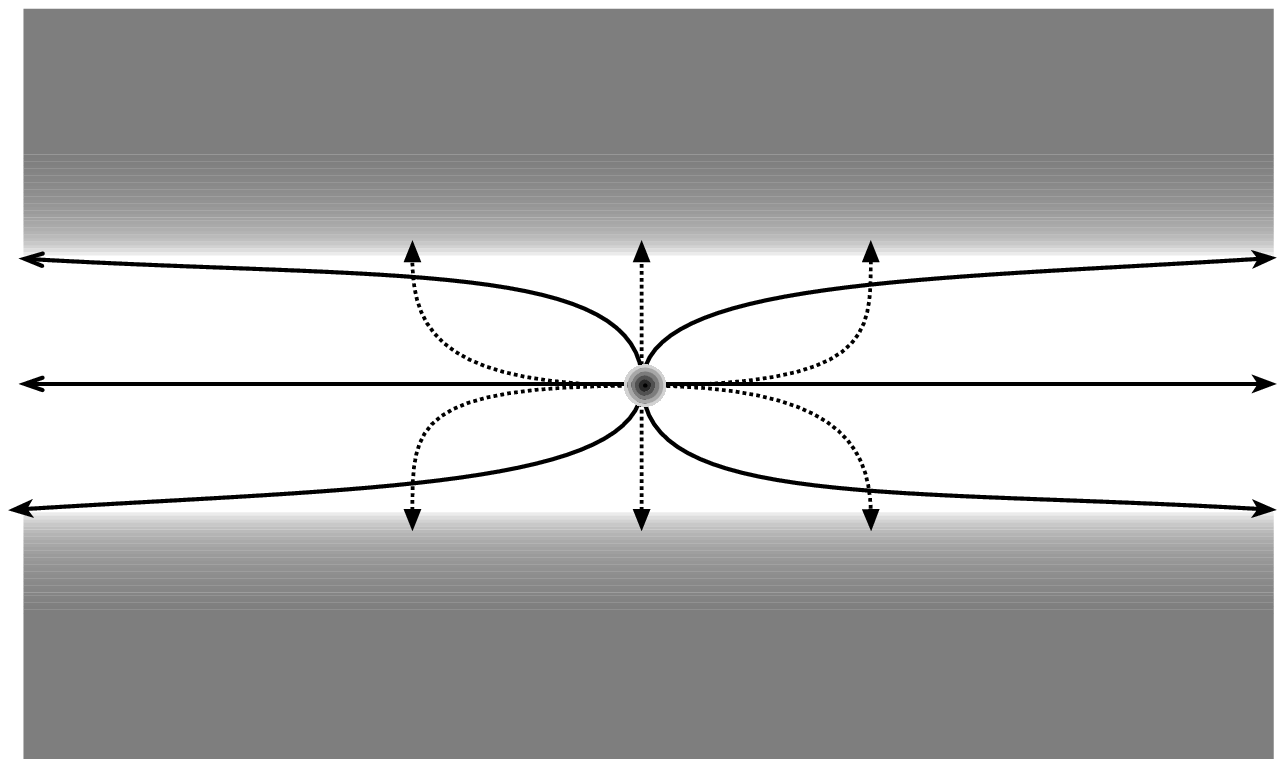}
\caption{Electric (dotted) and magnetic (solid) field lines generated by a fictitious dyon
(electrically and magnetically charged particle) in the Josephson junction. In the dual
picture the roles of the electric and magnetic fields are reversed.}
 \label{josephson}
 \end{center}
  \end{figure}

Let us consider the Abelian Higgs model in 3+1 dimensions, described by the
action 
\be
S = \int d^4x~\left\{
-\frac{1}{4} F_{\mu \nu} F^{\mu \nu} 
+
\left| \left( \partial_\mu + igA_\mu\right) \Phi\right|^2
-V(\Phi) \right\},
\label{lagr} \ee
with $F_{\mu\nu}=\partial_\mu A_\nu-\partial_\nu A_\mu$.
In a simple configuration that could localize the gauge field on a brane,
the scalar field $\Phi$  has
a zero expectation value inside the
brane and a constant non-zero value in the bulk. 
For our discussion we shall assume that $\Phi=0$ for $|z|< d/2$ and
$\Phi=\rho \exp(i\omega)$ for $|z| \geq d/2$. As a result, the gauge 
field becomes massive in the bulk through the Higgs mechanism, while
it remains massless inside the brane. 

The Abelian Higgs model in the spontaneously broken phase is
equivalent to the Ginzburg-Landau theory of superconductivity. 
A well known property of superconductors is the existence of 
frictionless currents that can flow without a potential. 
For the tree-level Lagrangian of eq.~(\ref{lagr}) the current density is
given by 
\be 
J_{tr}^\mu= -2 g \rho^2 \partial^\mu \omega - 2 g^2 \rho^2 A^\mu.
\label{current} \ee
The phase $\omega$ is unobservable for a bulk superconductor,
as it can be eliminated through a gauge transformation.
The superconducting currents flow near the surface of the material
and expel magnetic fields from it. This property, the Meissner effect,
can be understood also as a consequence of the photon mass $\sqrt{2}g \rho$.
The magnetic field decays over a distance $\lx \sim (g \rho)^{-1}$. Deep
inside the superconductor, the gauge field is zero and the current
of eq.~(\ref{current}) vanishes.

The resulting configuration (Fig. \ref{josephson})corresponds to a 
Josephson junction: two superconducting regions
separated by a thin layer of non-superconducting material. 
The phase $\omega$ cannot be eliminated completely in
this case. More specifically, 
we can define the gauge-independent phase difference between 
two points \cite{joseph}
\be
\dth_{P_1 P_2} 
= \omega(P_2)-\omega(P_1) - g\int_{P_1}^{P_2} \vec{A}\cdot d\vec{l}.
\label{dth} \ee
The gauge-dependent phases 
$\omega(P_1)$ and $\omega(P_2)$ can be eliminated for convenience through
an appropriate gauge transformation. The phase difference $\dth$
between two
points on either side of the junction can be non-zero \cite{joseph}.

As a result of the presence of a phase difference, a superconducting 
current, the Josephson current, flows across a junction.
This can be shown on general grounds \cite{weinberg} by noticing that,
beyond tree level, the effective action of the system 
depends on $\dth$ through quantum effects (tunnelling of charges
across the junction). 
The current can be obtained from the effective matter 
action through differentiation with respect to the gauge field.
The dependence of the action on a phase difference given by 
eq.~(\ref{dth}) immediately leads to the presence of a current. This 
tunnelling current does not require 
a potential difference between the two superconducting regions in order
to flow.

We can change each of the phases $\omega(P_1)$, $\omega(P_2)$ by a multiple
of $2\pi$ without altering its physical significance. This means that the
effective action and the Josephson current must be periodic
functions of $\dth$. We parametrize the current density 
as \cite{joseph,weinberg}
\be
J^3(\dth) = J^3_{\max} \sin(\dth),
\label{jz} \ee 
where $J^3_{\max}$ is its maximum value.
We emphasize that $J^3$ is a tunnelling current. 
(The classical current of eq.~(\ref{current}) is zero in the brane.) 
The maximum value $J^3_{\max}$ can be taken as a phenomenological parameter that 
can be very small 
in units of the typical mass scale of the potential
$V(\phi)$ in eq.~(\ref{lagr}). 

We can now consider the behavior of the electromagnetic field inside
the brane. We employ the Maxwell equations 
in the presence of an external current density given by eq.~(\ref{jz}).
They correspond to the equations of motion
derived from the tree-level Lagrangian of eq.~(\ref{lagr}) with $\phi=0$ and
an external current. 
The presence of the bulk superconductors imposes certain 
conditions on the solutions of these equations. 
The electric field parallel to a conductor is zero near its surface.
As we discussed in the previous section based on the arguments of \cite{giamisha},
this means that electric field lines
must end perpendicularly to the boundary of the Josephson junction (Fig. \ref{josephson}).
For a point charge, the electric field dies off within a distance
$ \sim d$ in the $x,y$-directions. 
As we are interested in the low-energy behavior of the 
system, we do not consider configurations with variations of the fields
at short distances. This means that we can approximate 
$E_x=-{\partial A^0}/{\partial x}-{\partial A^1}/{\partial t}$ and
$E_y=-{\partial A^0}/{\partial y}-{\partial A^2}/{\partial t}$
as zero 
and assume that $E_z=-{\partial A^3}/{\partial t}$
is independent of $z$ inside the brane.
The magnetic field has a continuous $z$-component at the surface of 
a conductor. As it is zero inside the superconductor, we 
assume that $B_z=-{\partial A^1}/{\partial y}+{\partial A^2}/{\partial x}$
vanishes everywhere. The components $B_x={\partial A^3}/{\partial y}$,
$B_y=-{\partial A^3}/{\partial x}$
are non-zero in the brane and vanish exponentially within a distance 
$\lx \sim (g \rho)^{-1}$ in the bulk. The magnetic field lines are
localized inside the brane (Fig. \ref{josephson}).

It is clear that the only unconstrained component of the gauge field is $A^3$, whose 
value is related to the gauge-invariant
phase difference $\dth$ across the brane.
From eq.~(\ref{dth}) with $\omega=0$ we obtain
\be
\dth = ~-g\int_{P_1}^{P_2} \vec{A}\cdot d\vec{l}
~\simeq~ \dth_{P_1P_2} = -g d A^3(P)
\label{dthabcd} \ee
Here $P_1$, $P_2$ are points opposite each other on either side of the brane and 
$A^3(P)$ the value of the gauge field inside the brane. 
The equation of motion of of $\dth$ can be derived through an elementary use of Maxwell's equations
\cite{joseph,tetradis}. It is
\be
\partial^i\partial_i \dthp + g d J^3_{\max} \sin ( \dth) = 0,
\label{eomdthsine} \ee
where $i=0,1,2.$ 
We conclude that there is one light mode on the brane that obeys the 
sine-Gordon equation. This mode corresponds to the third component of
the gauge field or the phase difference between the condensates on either
side of the brane: $\dth = -g d A^3$.
If we consider weak fields $(\dth \ll 1)$, 
we can approximate  
eq.~(\ref{eomdthsine}) as 
\be
\left[ \partial^i\partial_i + m^2 \right] \dth =0,
\label{eomdth} \ee
with $m^2=g dJ^3_{\max}$. 
In realistic Josephson junctions 
eq.~(\ref{eomdth}) implies the presence of a Meissner effect 
even in the non-superconducting material \cite{joseph}. 
Applied electromagnetic fields
decay over a distance 
$\sim \left(g d J^3_{\max}\right)^{-1/2}$. This phenomenon has
been observed experimentally. The decay length in the junction can
be orders of magnitude larger than the decay length in the
superconductor.
Also solitonic configurations can appear, 
corresponding to 
solutions of eq.~(\ref{eomdthsine}) \cite{joseph}.

We can conclude that the effective 
(2+1)-dimensional theory includes only one light physical degree of freedom ($A^3$ or $\dth$).
The non-vanishing components of the electromagnetic field are 
\be
B_x=\frac{\partial A^3}{\partial y}
~~~~~
B_y=-\frac{\partial A^3}{\partial x}
~~~~~
E_z=-\frac{\partial A^3}{\partial t},
\label{sbmub} \ee
where we have assumed no $z$-dependence.
However, we would like to have non-vanishing $E_x$, $E_y$, $B_z$ in order to 
generate an effective (2+1)-dimensional theory of electric charges. 
A possible remedy for this situation is provided by the suggestion
of ref.~\cite{giamisha}. 
The material in the bulk must be a {\it dual}
superconductor \cite{dualsup}. 
In other words, there must be a condensate of magnetic charge in
the bulk.

\subsection{The dual Josephson junction}

It is believed that dual superconductivity is realized in the 
confining phase of gauge theories. The particular implementation 
of ref.~\cite{giamisha} employs
an $SU(2)$ gauge theory coupled to a scalar field in the
adjoint representation (the Georgi-Glashow model) 
\cite{georgi}. Inside the brane 
the $SU(2)$ symmetry is broken down to $U(1)$ through a 
non-zero expectation value of the scalar field. The low-energy theory is in
the Coulomb phase and a  massless photon 
should emerge.
In the bulk the scalar field has a zero expectation value and
the theory is in the confining phase. 
All excitations are very massive, and this prevents the photon that is 
localized on the brane from entering the bulk.

In our discussion here we shall use only the main elements
of the above picture.
We consider electromagnetism in the presence of $U(1)$ magnetic charge. 
We assume that a 
magnetic condensate forms in the bulk, with the appearance of
frictionless currents. In the absence of electric charge, we
can use a phenomenological description
\be
S = \int d^4x~\left\{
-\frac{1}{4} \Ft_{\mu \nu} \Ft^{\mu \nu} 
+
\left| \left( \partial_\mu + ig_m C_\mu\right) \psi\right|^2
-V(\psi) \right\}.
\label{lagrd} \ee
The dual gauge field $C^\mu$ is defined through the duality 
\cite{zwanziger,report}
\be
F^{\mu \nu} \to \Ft^{\mu \nu} =\frac{1}{2} \epsilon^{\mu \nu \lambda \sigma} 
F_{\lx \sigma}= \partial^\mu C^\nu-\partial^\nu C^\mu, 
\label{dualfield} \ee
and $g_m$ is the magnetic charge.
We emphasize at this point that, in the presence of a magnetic
current, the field $F^{\mu\nu}$ does not have a simple description in terms of a gauge field
$A^\mu$ \cite{report}. Only the dual field $\Ft^{\mu \nu}$ can be expressed simply through
$C^\mu$. 
A frictionless magnetic current flows near the surface of the
regions of non-zero expectation value for
the magnetic condensate
 $\psi=\sigma \exp(i\chi)$.
At tree level it is given by 
\be 
\Jt_{tr}^\mu= - 2 g_m \sigma^2 \partial^\mu \chi - 2 g_m^2 \sigma^2 C^\mu.
\label{currentd} \ee
A dual Meissner effect prevents the electric field from entering 
the regions with $\psi\not= 0$. 

The system of Fig. \ref{josephson} can be viewed now as a dual Josephson junction 
with a tunnelling magnetic current $\Jt^3$ flowing across the brane. 
It is clear from eq.~(\ref{currentd})
that the gauge-invariant definition of the phase
difference $\dch$ between the two sides of the brane must involve the
dual field $C^\mu$. 
Repeating the arguments that led to eq.~(\ref{jz}) we find
\be
\Jt^3(\dch) = \Jt^3_{\max} \sin(\dch).
\label{jzd} \ee 
We emphasize at this point 
that the presence of a current is independent of the detailed form of the
Lagrangian of the system. It is a consequence only of our assumption
that a condensate exists on either side of the brane 
\cite{weinberg}. 

We turn next to the gauge field localized
on the brane. 
The Maxwell equations read 
\begin{eqnarray}
\partial_\mu F^{\mu\nu} &=& 0,
\label{maxwell1} \\
{\partial_\mu} \Ft^{\mu\nu} &=& \Jt^\nu.
\label{maxwell2} \end{eqnarray}
In the second equation we have included 
the tunnelling magnetic current $\Jt^3$.
We must also take into account the constraints on the
electromagnetic field arising from the presence of the dual
superconducting phase in the bulk. The arguments we gave in the case of the
standard Josephson junction can be repeated with the exchange of
the role of electric and magnetic fields (Fig. \ref{josephson}).

The Maxwell equations can be solved easily in terms of the dual 
gauge field $C^\mu$
defined in eq.~(\ref{dualfield}). 
The non-vanishing components of the electromagnetic field are 
\be
E_x=-\frac{\partial C^3}{\partial y}
~~~~~
E_y=\frac{\partial C^3}{\partial x}
~~~~~
B_z=-\frac{\partial C^3}{\partial t},
\label{bmub} \ee
where we have assumed no $z$-dependence.
The reasoning that led to eq.~(\ref{eomdth}) now gives
\be
\partial^i\partial_i \left( \dch \right) + g_m d \Jt^3_{\max}
\sin \left( \dch \right) = 0.
\label{eomdchsine} \ee
For $\dch \ll 1$ we obtain 
\be
\left[ \partial^i\partial_i + \mt^2 \right] \dch =0,
\label{eomdch} \ee
with $\mt^2=g_m d \Jt^3_{\max}$.
The massive mode $\dch$
can be identified with the third component of
the dual field: $\dch=-g_m d\, C^3$.

In summary, the following picture emerges: A (2+1)-dimensional low-energy
theory appears on the brane. It involves one physical degree of freedom
that can be identifed with the dual gauge 
field $C^3$. The component of the electromagnetic field 
$E_x$, $E_y$, $B_z$ are given by the simple expressions~(\ref{bmub}).
The field $C^3$ is massive, with a mass $\mt$ that can
be very small in units of the typical scale of the theory
in the bulk. As a result, the electromagnetic field 
has a finite correlation length.
This can be seen
by simply taking $y$, $x$, $t$-derivatives of
eq.~(\ref{eomdch}) and remembering that $\dch=-g_m d \,C^3$.
Indirect experimental support of this picture comes from
the observation of a Meissner effect in standard Josephson
junctions. 

We expect the above conclusions to remain valid in a theory that includes
electric charges on the brane. As we mentioned earlier, the 
Josephson effect is an immediate consequence of the 
presence of charged condensates and does
not depend on the details of the underlying theory \cite{weinberg}.
Therefore, the complications encountered in constructing a
consistent theory of electric and magnetic charges are not expected
to lead to significant modifications of our arguments. 
We mention that, in a consistent theory, electric and magnetic 
charges must satisfy Dirac's quantization condition
\cite{dirac}
\be
g_e g_m = 2\pi n.
\label{quant} \ee

An interesting solution of the sine-Gordon equation~(\ref{eomdchsine}) 
is given by \cite{joseph}
\be
\dch = 2 \sin^{-1} {\rm sech} \left[ \mt \left(x -x_0 \right)\right].
\label{fluxon} \ee
It corresponds to a defect localized near the line $x=x_0$. The electric
field $E_y$ is non-zero near $x_0$ and vanishes
at distances $\gta \mt^{-1}$ away from it.  
The phase $\dch$ changes by $2\pi$ as $x$ goes from $-\infty$ to 
$\infty$. It is easy to see that the defect carries
unit electric flux $2\pi/g_m = g_e$, for $n=1$ in eq.~(\ref{quant}). 
The energy per unit length of the 
defect is $\sim \left( \Jt^3_{\max}/g_m^3d\right)^{1/2}$ \cite{joseph}.
Lines with larger electric flux correspond to solutions for which
$\dch$ varies by multiples of $2\pi$. It is natural to expect that flux-carrying 
lines of this type connect opposite electric charges on the brane.

There is a close similarity between the bulk and the brane of width $d$. 
On the brane the electromagnetic field is massive and defects exist that
carry electric flux, similarly to the behavior in the bulk. 
The most consistent interpretation of the emerging (2+1)-dimensional theory is 
that it displays confinement with a linear potential, but with a typical 
scale much smaller than the scale characterizing the theory in the bulk.
This is in agreement with 
the experimental studies of standard Josephson junctions.
In that case the system behaves as if the
superconducting properties extend 
over the whole structure including the barrier \cite{joseph}.
In a certain sense, this is caused by 
the electric condensate penetrating the barrier instead of ending
abruptly at the surface. 
For the dual picture that we are considering, we expect the magnetic
condensate to behave in an analogous way. The implication is 
that dual superconducting behavior, and therefore confinement, 
must be present inside the brane as well.

\subsection{Compact QED in 2+1 dimensions}

The $U(1)$ gauge theory that emerges at low energies through the
localization mechanism discussed in section \ref{localization} is compact. This is a 
consequence of the fact that the $U(1)$ symmetry is embedded in the non-Abelian
$SU(2)$ group. The low-energy theory has strong similarities with the Georgi-Glashow model
\cite{georgi} in 2+1 dimensions.
The unbroken $U(1)$ gauge symmetry in this model again results from the breaking of 
$SU(2)$. The theory contains instantons, which correspond to the monopoles of
the (3+1)-dimensional theory when the Euclidean time and the third spatial dimension
are interchanged. The smallest magnetic charge is $g_m=4\pi/g$, where $g$ is the gauge coupling
constant.
The unit electric charge in the Georgi-Glashow model is
$g_e=g/2$, so that $g_e$, $g_m$ satisfy
eq. (\ref{quant}) with $n=1$. 

Polyakov \cite{polyakov} has demonstrated that the presence of instantons
destroys the Coulomb phase, moving the theory to a confining one. The sector with $n=1$ in
eq. (\ref{quant}) dominates the dynamics.
The confinement scale is
exponentially suppressed by the action of the instantons. 
This conclusion is in agreement with the physics of the Josephson 
junction. In the Josephson picture the confinement is a consequence of the tunnelling 
monopoles of the (3+1)-dimensional theory. It seems natural to identify them with the 
instantons of the (2+1)-dimensional theory. In the following section we shall make this
relation concrete.

Before discussing the details of this connection, we point out some other striking similarities
between the two pictures. 
In the (2+1)-dimensional theory, 
the confining dynamics is dominated by the presence of widely separated instantons.
The relevant 
partition function is very similar to that of the Coulomb gas. 
In this work we are interested in the case of a large Higgs mass
($m_H \gg m_W$ at the minimum $\Phi_0$ of the Higgs potential)\footnote{
For $m_H \gg m_W$ the boundaries of the brane are very sharp, in agreement with our assumptions throughout
the paper.}.
The partition function of the (2+1)-dimensional theory can be cast in the form \cite{polyakov,dietz}
\be
Z \sim \int {\cal D}\chit 
\exp \left(
-\frac{g^2}{32\pi^2}\int d^3x \, \left[
\partial^\mu \chit \partial_\mu \chit + 2 \Mt^2 \cos\left(\chit \right)
\right] \right),
\label{coulomb} \ee
with 
$\Mt \sim \exp(-S_0)$, where $S_0$ is the instanton action $S_0=4\pi\ex\Phi_0/g$, with $\ex\simeq 1.787$.
The field $\chit$ is the dual photon and corresponds to the 
single physical degree of freedom of the (2+1)-dimensional gauge theory. Its 
equation of motion is given by (\ref{eomdchsine}), with  
$\Jt^3_{\max}\sim \exp(-S_0)$. It is clear that the field $\chit$ of the
(2+1)-dimensional theory can be identified with the field
$C^3\sim \dch$ in the picture of the Josephson junction. 

It is obvious from the above that the physics of the localized gauge theory can be described equivalently 
in two different ways: a) through the picture of the Josephson junction, with the 
tunnelling monopoles playing a major role; b) in purely (2+1)-dimensional terms, through the compact 
$U(1)$ gauge theory, in which the instantons are the dominant entities.
In the following, we demonstrate the equivalence of the two pictures employing topological arguments.

\section{Description in Terms of Topological Entities}

 We now wish to establish a connection between the long-distance confining dynamics of the
 $U(1)$ theory on the brane and the confinement in the $(2+1)$-dimensional 
 $SU(2)$ Higgs theory.  Qualitatively, it is clear that this connection should come from the fact that in 
both cases there is a $U(1)$ symmetry embedded in $SU(2)$ above a certain scale.  
However, the difficulty in applying the experience from the  
$(2+1)$-dimensional case arises because the 
$U(1)$ theory gets directly embedded in a $(3+1)$-dimensional $SU(2)$. This, although
Higgsed on the brane,  is never in a Higgs phase in the $(2+1)$-dimensional sense. 
In other words, the localization of the $(2+1)$-dimensional $U(1)$ is a result of the interplay
between the bulk confinement and the brane Higgs effect. The two effects cannot be decoupled.  
There is no intermediate window of scales within which one could decouple the confining bulk physics, 
and in the same time ignore the brane Higgs effect, in such a way that the effective 
theory in this interval  can be regarded  as a
$(2+1)$-dimensional  $SU(2)$.    

In order to circumvent this complication we shall employ topological arguments, 
which will allow us to trace the origin of the instantons in the effective low-energy theory on the brane 
in terms of higher-dimensional topological entities. This will allow us to bypass the full complicated
confining dynamics in the bulk.  
The method that we shall discuss is useful more generally for visualizing the topological structure 
of theories in arbitrary dimensions in terms of topological entities of a  higher-dimensional theory, 
even if the latter is a pure mathematical extrapolation.    

The connection between the topological defects in a $(D+1)$-dimensional theory with the instantons 
in $D$-dimensions is well known. The important point of this connection is that,  
by viewing the Euclidean time 
coordinate in $D$ dimensions as the $D$-th space coordinate in $D+1$ dimensions,  
the $D$-dimensional instanton solution  
becomes a topological defect in $D+1$ dimensions. 

In the same time instantons in $D$ dimensions describe transitions between vacua with different 
topological winding numbers. 
The transition between different vacua in $D$ dimensions 
can be  given an explicit topological meaning in terms of the motion of a
soliton in $D+1$ dimensions as follows: 
Imagine that the spatial part of the $D$-dimensional theory is a $D-1$ sphere of radius $R$, embedded 
in a fictitious $D$-dimensional space. 
The fields of the lower-dimensional theory $\Phi(x)$, where $x$ denotes the
world-volume coordinates of the sphere,
can be viewed as expectation values of 
higher-dimensional fields $\Phi(x,y)$ on the sphere ($y=R$). 
A topological vacuum
of the $D$-dimensional theory with non-zero winding number simply corresponds to  
a configuration in which a monopole is placed at the center of the sphere in $D$ space dimensions. 
The $D$-dimensional instanton driving the transition  between different vacua 
corresponds to the motion of the monopole across the sphere in the $D+1$-dimensional picture. 
Changing the winding number in the first picture 
is equivalent to changing the monopole number enclosed by the sphere in the second picture.
This connection becomes much clearer in the specific examples we consider in the following.

\subsection{Instantons in Two Dimensions as Tunnelling Vortices}

Consider the $(1+1)$-dimensional Abelian Higgs model
\be
S = \int d^2x~\left\{
-\frac{1}{4} F_{\mu \nu} F^{\mu \nu} 
+
\left|  {\cal D}_\mu  \Phi\right|^2
-\lx \left(\left|\Phi\right|^2-{v^2}\right)^2 \right\},
\label{2dhiggs} \ee
 where $\Phi \equiv \rho \exp({i\omega})$ is a complex scalar Higgs field.  The minima of the
potential form a circle with radius $\rho = v$.  
Let us imagine that the spatial dimension is also a compact circle of some radius $R$ much larger than
the inverse Higgs mass. 
The possible vacua of this theory  can be characterized by a non-trivial topological winding number  
 \begin{equation}
 n = {\frac{1}{2\pi}} \int  d\phi \,\left( \partial \omega/\partial \phi\right)=
\frac{\omega(2\pi)-\omega(0)}{2\pi}.
 \label{numbern}
 \end{equation} 
The explicit form of the Higgs field is 
$\Phi_{vac} = v \exp \left({i{n}\phi}\right),$
with the coordinate $\phi$ changing from 0 to $2\pi$ around the circle. The configuration of the 
gauge field corresponds to a pure gauge. 
Similar vacua exist also in the non-compact case, with the fields at the points $x=-\infty$ and $x=\infty$
differing by a large gauge transformation.

In this theory there are also
instantons, that correspond to transitions between vacua with different $n$ 
\cite{rajaraman}. 
If the spatial dimension is non-compact, 
an instanton is given by a vortex configuration \cite{vortex}
in the two-dimensional space that results from switching to imaginary time $t_e$.
Its explicit form is
\begin{eqnarray}
\Phi_{inst} & = & f(r_e) \exp \left[{in\phi_e}\right]
\nonumber \\ 
r_e^2&\equiv&x^2+t^2_e
\nonumber \\
\phi_e&\equiv&\arctan\left( t_e/x\right).
\label{vvortex} \end{eqnarray}
Here $f(r)$ is the Nielsen-Olesen function that satisfies $f(0) =0,~f(\infty) = v$ \cite{vortex}.
The gauge field corresponds to a pure gauge.    

For a compact spatial dimension the instanton transition can be  
visualized as follows: Imagine that   
the circle of radius $R$ on which our model lives is embedded in a two-dimensional plane.   
The $(1+1)$-dimensional fields can be viewed as 
being determined by their higher-dimensional extensions evaluated on the circle.  
The vacua with winding number $n$ 
on the circle correspond to configurations in which there are $n$ vortices enclosed by the circle.
This becomes obvious if we place a vortex with winding number $n$ at the center. This configuration on
the circle
is described by the first relation of 
(\ref{vvortex}) with the replacements $\phi_e \to \phi$, $r_e \to R$, where
now $\phi$ is the angular coordinate.
Obviously, the integral (\ref{numbern}) is then equal to $n$. 
The instantons that change this winding number correspond to the change of the 
vortex number inside the circle in the (2+1)-dimensional picture!  

\begin{figure}[t]
\begin{center}
\includegraphics[clip,width=0.8\linewidth]{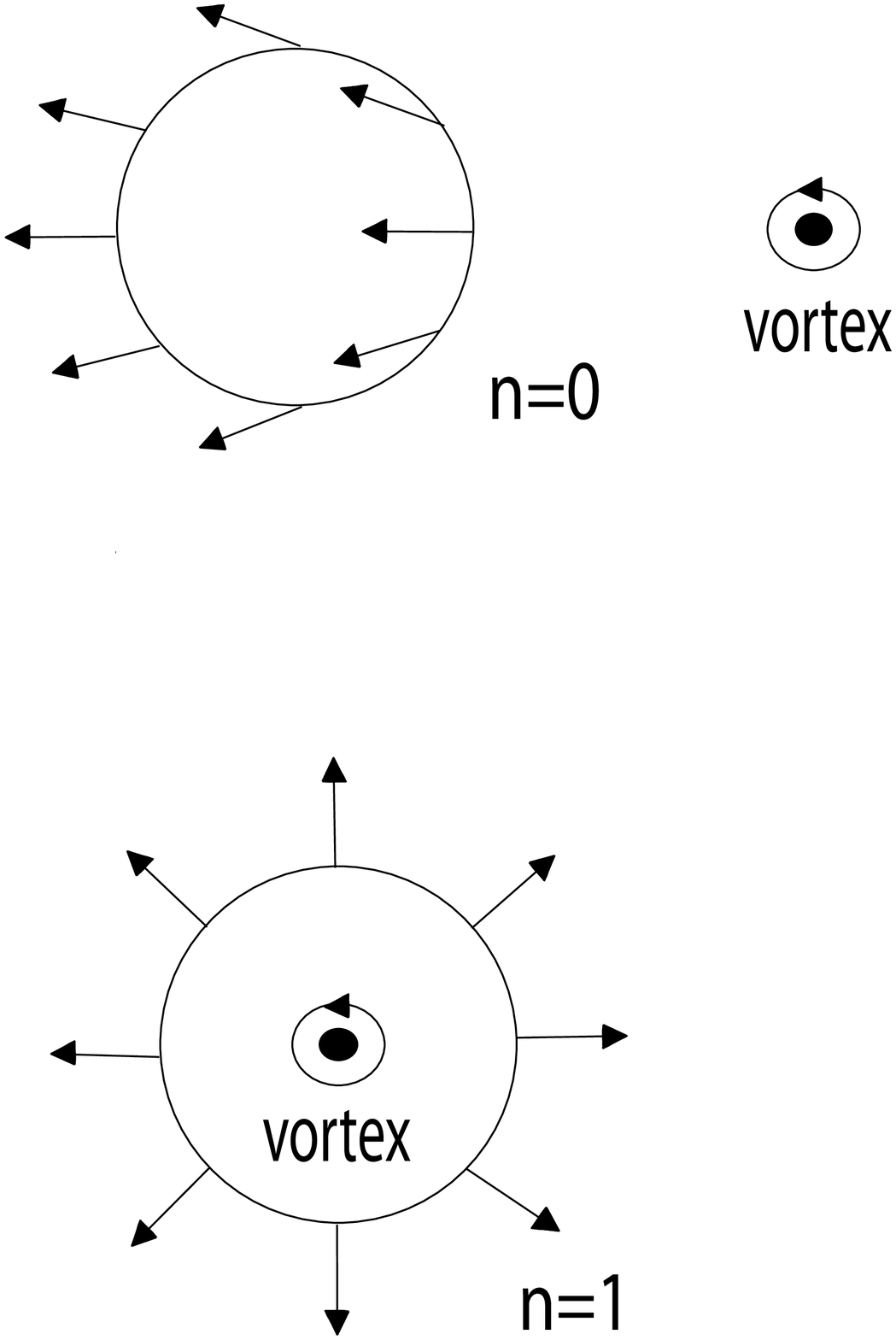}
\caption{A process that changes the winding number of the vacuum by one unit in the (1+1)-dimensional 
Abelian Higgs model.}
 \label{monopole}
 \end{center}
  \end{figure}

For example, the transition from a trivial vacuum to one with winding number $n$ is equivalent to a
vortex with number $n$ entering the circle of radius $R$. This process is depicted in Fig. \ref{monopole}
for $n=1$. The black dot denotes a vortex, initially located outside the circle. The arrows indicate
the orientation of the Higgs field in internal space. They are taken to have equal length, as the
magnitude of the Higgs field is $v$ far from the vortex center. It is apparent that initially the
vacuum winding number for the (1+1)-dimensional theory on the circle is zero. When the vortex
moves inside the circle the winding number increases by one unit.

\subsection{Instantons in Three Dimensions as Tunnelling Monopoles}

The discussion of the previous subsection can be generalized to higher dimensions. 
Consider the $(2+1)$-dimensional $SU(2)$ Higgs model
\be
S = \int d^3 x~\left\{
-\frac{1}{4} F^a_{\mu \nu} F_a^{\mu \nu} 
+\mathcal{D}_{\mu} \Phi^a\mathcal{D}^{\mu}\Phi_a  -  \lx\left( \Phi^a\Phi_a  -  v^2\right)^2 
\right\}
\label{3dhiggs}
\end{equation}  
with the Higgs field in the adjoint represenation $a=1,2,3$. 
There exist various vacua that  
correspond to topologically non-trivial configurations. 
If the two-dimensional space in non-compact, the simplest such configuration is given by 
\begin{equation}
\label{widing2}
\Phi \, = \, v\, (\cos\phi \sin\theta(\rho), \sin\phi \sin\theta(\rho), \cos \theta(\rho)), 
\end{equation}
in polar coordinates $\phi, \rho$.  
The function
$\theta(\rho)$ interpolates between $0$ and $\pi$ as $\rho$  goes from $0$ to $\infty$.  
The corresponding topological number is 
\begin{equation}
\label{inv}
n=\frac{1}{8\pi v^3}
\int d^2x \, \epsilon_{abc}\, \epsilon^{0\mu\nu }\,  \partial_{\mu} \Phi^a\partial_{\nu} \Phi^b \, \Phi^c=1. 
\end{equation}   
The two-dimensional space can be compactified to a two-sphere of radius 
$R$ by identifying the points at infinity ($\rho \to \infty$). We can
imagine this sphere being embedded in a fictitious three-dimensional  
space. 
The configuration of eq. (\ref{3dhiggs}) 
would originate from a monopole located at the center of the sphere.  
The (2+1)-dimensional instanton then would describe a process that moves a monopole through the 
two-sphere. 
During this process the winding number (\ref{inv}) changes by one unit.  
This demonstrates our main point: 
The (2+1)-dimensional instanton can be understood as a process that takes the monopole 
across the space from an imaginary third dimension. 
  
The application to the problem of gauge field localization is
straightforward.  We consider a 
(3+1)-dimensional $SU(2)$ model 
with an adjoint Higgs
\be
S = \int d^4 x~\left\{
-\frac{1}{4} F^a_{\mu \nu} F_a^{\mu \nu} 
+\mathcal{D}_{\mu} \Phi^a\mathcal{D}^{\mu}\Phi_a  
-  V(\Phi) \right\}.
\label{4dhiggs}
\end{equation}  
The potential $V(\Phi)$ has two degenerate minima, one with 
$\Phi=0$ and the other with $\Phi^a\Phi_a  =  M^2$. An example of such a potential
is given by (\ref{potu2}). 
The first vacuum is confining at some scale $\Lambda$ and there is a mass gap. 
There are no massless states there.  In the second vacuum $SU(2)$ is Higgsed down to 
$U(1)$ and there is a massless photon. The two phases can coexist, separated by walls of 
thickness $\sim M^{-1}$ and tension $\sim M^3$.  We assume that $M \gg \Lambda$. We consider 
a layer of the $\Phi\neq 0$ phase  sandwiched between the $\Phi=0$ phases. 
We assume that the thickness of the layer is $d \gg M^{-1}$.   The walls then have an exponentially 
suppressed interaction and can be considered to be static during the time scales of interest. 

The important point is that configurations in this model, 
just like the ones in the trully $(2+1)$-dimensional 
Higgs model, can be 
characterized by a topological winding number. An example is given by  
\begin{equation}
\label{number}
\Phi \, = \, f(z) \, (\cos\phi \sin\theta(\rho), \sin\phi \sin\theta(\rho), \cos \theta(\rho))
\end{equation}
where $z$ is the coordinate perpendicular to the layer, and $f(z)$ describes the 
parallel wall and antiwall: It vanishes for $z\to \infty$ and $-\infty$ and takes the value
$M$ within
a layer of width $d$ around $z=0$.  The corresponding winding number is 
 \begin{equation}
\label{inv3}
n= \frac{1}{8 \pi v^3}
\int d^3x \, \epsilon_{abc}\, \epsilon^{30\mu\nu} \partial_{\mu} \Phi^a\partial_{\nu} \Phi^b\,  \Phi^c.
\end{equation} 
This winding number can be changed by taking a monopole through the layer. The instanton of
the $(2+1)$-dimensional theory describes 
this process.  
As the mass of the monopole in the layer is given by the expectation value of $\Phi = M$, 
in the limit $M  \gg  \Lambda \gg  d^{-1}$ the process must be exponentially suppressed by the factor 
$\exp({-Md})$.  The reason is that $Md$ sets the barrier that the  monopoles have to tunnel through.  
In the opposite case $M \sim d^{-1} \gg \Lambda$ the suppression factor should be 
$\exp({-M\Lambda})$.  

\subsection{Form of the flux tubes}

\begin{figure}[t]
\begin{center}
\includegraphics[clip,width=0.9\linewidth]{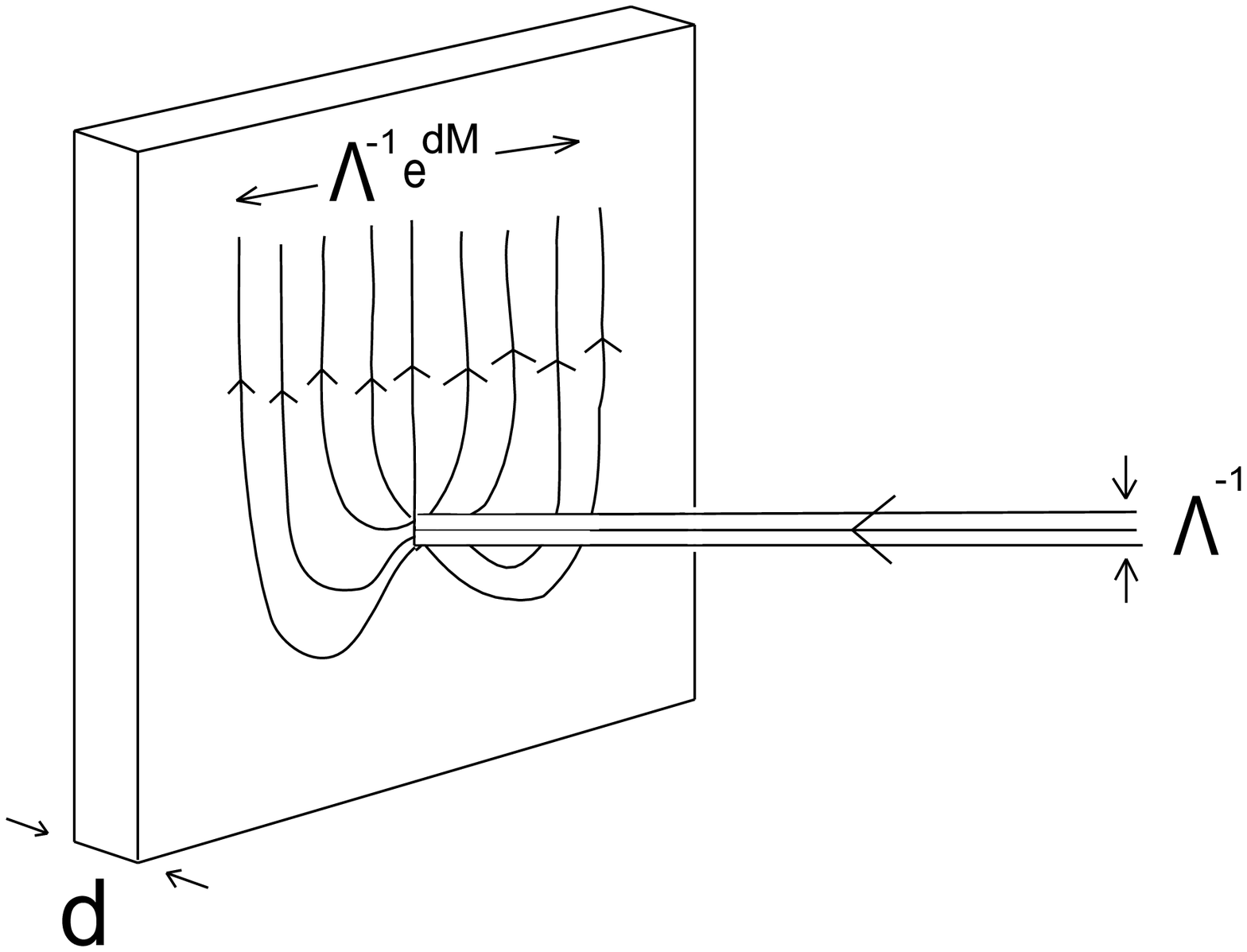}
\caption{Spreading of a flux tube (``open string'') on a layer (``D-brane'').}
 \label{flux}
 \end{center}
  \end{figure}

As we discussed in previous sections, the resulting low-energy theory is strongly confining in
the bulk, while it displays weak confinement of electric charges on the brane. 
In Fig. \ref{flux} we depict the different form of a flux tube in the bulk and on the brane. 
The configuration of Fig. \ref{flux} is generated by an electric charge located in the bulk and
another one on the brane. The electric flux lines form a tight tube in the bulk, of width 
$\sim 1/\Lambda$, where $\Lambda$ is the bulk confinement scale. 
On the brane the confinement scale is exponentially suppressed, so that the width of the
flux tube increases dramatically. On the other hand, no flux lines escape to infinity. They all
start and finish on the charges. This configuration has a very strong similarity to an
open string with its end attached on a D-brane in string theory.

\section{Four dimensions and Phenomenology}

\subsection{The Case of (3+1)-Dimensional Electrodynamics}

As we have seen, the effective $(2+1)$-dimensional electrodynamics becomes  
confining at exponentially large distances.  This happens because the UV-completing  $SU(2)$ physics 
does not decouple entirely, but leaves an imprint on the IR physics. This imprint, although
parametrically negligible at short length scales, becomes dominant at exponentially large distances 
and makes the $U(1)$ theory confining.

The natural question to ask is whether this effect is exceptional  to  $2+1$ dimensions, or whether it 
also takes place for the localization of the photon onto a $(3+1)$-dimensional brane in higher dimensions. 
The standard intuition in 3+1 dimensions tells us that the latter scenario is impossible, 
because, from what is known, the IR behavior of the pure $U(1)$-theory in $3+1$ dimensions is insensitive to 
the UV completion. 
However, as we are dealing with an extra-dimensional completion of the theory, 
the story may be much less straightforward. 

There are two possible cases:
\begin{itemize}
\item  
The first possibility is that the theory  
below a certain scale reduces to simple $U(1)$ electrodynamics with a massless photon with 
two polarizations, and no extra light degrees of freedom. In this case, we can make a strong 
argument that the $(3+1)$-dimensional localization is very different from the analogous one in $2+1$ 
dimensions. Also, 
confinement can take place in the latter case but not in the former.   

The argument is based on counting and matching the number of degrees of freedom for the massless 
$U(1)$ theory and 
the massive confining one.  
Consider the $(2+1)$-dimensional case first. 
As explained above, the theory includes two mass scales. The first one is the scale $d^{-1}$ (width
of the brane)  below which the effective low-energy theory  
is that of a massless $(2+1)$-dimensional photon, with one propagating degree of freedom. 
The second is the scale $\exp({-Md})$, below which the theory confines and becomes one of composite 
massive glueballs, with the lowest one being a scalar.  The crucial point is that the degrees of freedom  
needed for describing the two theories can match.  Both a massive scalar glueball and a massless 
vector field carry the same number of propagating physical degrees of freedom.  In other words, 
the massless $U(1)$ theory  can smoothly flow into the theory of a massive scalar glueball, 
without any need of extra degrees of freedom.  Notice that this would be impossible if the mass 
of the photon were of the Higgs or Proca type, since in such a case an  
extra  massless degree of freedom (a  Goldstone-St\"uckelberg field) is necessary,
that would become the longitudinal 
polarization of the massive photon.
  
In $3+1$ dimensions such a matching of modes is impossible.  The massless photon 
carries two degrees of freedom, and this number cannot be matched with the number of degrees of freedom 
in any massive representation of the Lorentz group. Thus,  the massless  $(3+1)$-dimensional  
$U(1)$ theory cannot smoothly flow to a theory of massive glueballs in the far infrared without 
acquiring  some extra  degrees of freedom from outside.  
    
\item
The second possibility is that the additional light degrees of freedom needed for the infrared confinement 
are provided by the UV-completing extra-dimensional  physics.  Even though we do not know of
a specific implementation of such a scenario, 
we cannot rule out this possibility, especially if the extra dimensions have infinite 
volume and no mass gap.  

\end{itemize}

\subsection{Phenomenological Bound on the Photon Confinement Scale} 

The most obvious experimental signature for the scenario we considered
is that 
electrodynamics would become 
confining at exponentially large distances. It is important to point out that the phenomenological bounds 
on such a confinement scale would be much  more severe than the bounds on a photon mass of 
the Higgs or Proca type.   
Interestingly,  this bound  follows from the existence of a long-range galactic magnetic field.  
This would be screened by the magnetic condensate if the confinement scale 
of the photon were larger than the inverse galactic size, that is $10^{-27}$eV! The bound 
on the Higgs-type mass of the photon is much milder, only about $10^{-16}$eV 
\cite{massphoton}, because such a mass 
cannot screen the galactic magnetic field, but only the electric one, which is absent anyway.

\section{Conclusions}  
\label{conclusions}

This paper was devoted to the investigation of the duality between the gauge-field localization mechanism 
of \cite{giamisha} and the physics of the Josephson junction. 
It was suggested in \cite{tetradis} that, if the low-energy duality is complete, 
the (2+1)-dimensional electrodynamics must become confining at exponentially large distances. 
By invoking topological methods we identified explicitly the sources of this confinement. 
They are the 
instantons that correspond to the tunneling across the brane of the monopoles that are condensed in
the bulk. The resulting current is  
completely analogous to the Josephson current.

The IR dynamics of the
theory that arises through the localization mechanism of \cite{giamisha}
seems sensitive to the UV completion. In particular, complete decoupling of the UV does not take 
place, as there is always a tunnelling current that is crucial for the IR behavior. 
This current flows because of the presence of magnetic condensates of the (3+1)-dimensional $SU(2)$ theory 
on either side of the brane.
It has, therefore, a (3+1)-dimensional character. On the other hand, the effect is exponentially suppressed.
The long-distance confining dynamics of the
$U(1)$ theory on the brane is very similar to the confinement in the $(2+1)$-dimensional 
$SU(2)$ Higgs theory with the symmetry broken down to $U(1)$.
In both cases there is a $U(1)$ symmetry embedded in $SU(2)$ above a certain scale.  
The long-distance dynamics reflects this embedding, which results in the presence of tunnelling monopoles in
the first case and instantons in the second. 

The most exciting aspect of the scenario we considered is that it has 
experimental low-energy implications for a non-gravitational sector. 
It demonstrates that the presence of extra dimensions 
can be unveiled much below the energies at which particles
are released in the bulk. 
We discussed the possible generalization of the effect for
higher dimensions and the phenomenological bound on the infrared confinement of the photon. 
Interestingly, this bound is much more severe than the one on a photon mass of the Higgs or Proca type. 
It follows from the existence of the galactic magnetic field, which cannot constrain 
a Higgs-type photon mass very efficiently.

\section{Acknowledgments}

The work of G.D. is supported by a 
David and Lucile Packard Foundation Fellowship for Science and Engineering and the
NSF grant PHY-0245068. 
The work of N.T. is supported by the research
program ``Pythagoras II '' (grant
70-03-7992) of the Greek Ministry of National Education, partially funded by
the European Union, the research program Kapodistrias of the University of Athens, and
by the European Commission under the Research and Training Network contract MRTN-CT-2004-503369.
N.T. would like to thank the CERN Theory Institute for its hospitality during the time that 
this work was carried out.

\end{document}